\documentclass[pra,aps,10pt, twocolumn, floatfix]{revtex4-1} 

\usepackage {graphicx} 
\usepackage {amssymb} 
\usepackage {amsmath}

\begin{document}

\title{Coherently controlled generation of single-cycle terahertz pulses from 
thin layer of  nonlinear  medium with low-frequency resonances}

\author{R. M. Arkhipov}

\affiliation{St. Petersburg State University, Universitetskaya nab. 7/9, St. Petersburg 199034, Russia}
\affiliation{ITMO University, Kronverkskiy prospekt 49, 197101, St. Petersburg,  Russia}
\affiliation{Ioffe Institute, Politekhnicheskaya str. 26, St. Petersburg 194021, Russia}

\author{A.V. Pakhomov}

\affiliation{ITMO University, Kronverkskiy prospekt 49, 197101, St. Petersburg,  Russia}

\author{M.V. Arkhipov}

\affiliation{St. Petersburg State University, Universitetskaya nab. 7/9, St. Petersburg 199034, Russia}
\affiliation{ITMO University, Kronverkskiy prospekt 49, 197101, St. Petersburg,  Russia}

\author{A. Demircan, U. Morgner}

\affiliation{Institute of Quantum Optics, Leibniz University Hannover, Welfengarten 1, 30167 Hannover, Germany}
\affiliation{Cluster of Excellence PhoenixD (Photonics, Optics, and Engineering – Innovation Across Disciplines), Welfengarten 1, 30167 Hannover, Germany}

\author{N.N. Rosanov}

\affiliation{Vavilov State Optical Institute, Kadetskaya Liniya v.o. 14/2, St. Petersburg 199053, Russia}
\affiliation{ITMO University, Kronverkskiy prospekt 49, 197101, St. Petersburg,  Russia}
\affiliation{Ioffe Institute, Politekhnicheskaya str. 26, St. Petersburg 194021, Russia}

\author{I. Babushkin}
\affiliation{Institute of Quantum Optics, Leibniz University Hannover, Welfengarten 1, 30167 Hannover, Germany}
\affiliation{Cluster of Excellence PhoenixD (Photonics, Optics, and Engineering – Innovation Across Disciplines), Welfengarten 1, 30167 Hannover, Germany}
\affiliation{Max Born Institute, Max-Born-Strasse 2a, Berlin 10117, Germany}




\begin{abstract}
 We propose a novel scheme to generate single-cycle terahertz (THz) pulses via reflection of an optical femtosecond pulse train from a thin layer of nonlinear resonant medium. Our method is based on a coherent control of low-frequency oscillations and free induction decay in the medium. 
The specific single-cycle shape of generated THz pulses requires
a plane wavefront and detection in the near field.
Our theoretical results pave the way to a new, simple and high-efficiency way to generate single-cycle waveshape-tunable terahertz pulses.
\end{abstract}


\maketitle
 Generation of few-cycle pulses in the teraherz (THz) spectral range (0.1--30 THz) is nowadays of great interest due to the huge amount of their applications in science and technology \cite{Roskos, Reiman, Lepeshov}. Among them are ultrafast spectroscopy of semiconductors, dielectrics and  molecules with resonances in THz range, biological sensing, detection of dangerous substances, medicine etc.  
Present-day methods of THz generation are mainly based on frequency conversion of intense femtosecond optical pulses in nonlinear media. The mechanisms of generation are related to four-wave mixing \cite{Roskos, Reiman} or ionization-based nonlinearity  \cite{Kim1, Kress1, Babushkin1,Tcypkin}.  Conceptually similar idea is used when THz frequencies are emitted via optically excited photoconductive switchers, nowadays often enhanced by nanostructures \cite{Reiman,Lepeshov,Berry2,Gorodetsky}. In parallel, high efficient methods of THz pulse synthesis via optical rectification in different materials have been developed so far \cite{Palfalvi,Nugraha,Toth}.
In \cite{Arkhipov-OL} a novel way of unipolar half-cycle pulse production via reflection of single-cycle pulse from linear broad thin metallic or dielectric medium was proposed. This way allows generation either optical or THz subcyle wave forms.  However,  using single-cycle pump pulse is crucial point of the proposed scheme, which makes practical realization of the proposed method significantly difficult. Nevertheless, generation of tunable ultrashort THz waveforms with high efficiency still constitutes a severe problem. 

In this letter, we propose and study theoretically a new way to produce single-cycle THz pulses with tunable waveshape using a rather simple and compact setup. It significantly differs from recently proposed technique of half-cycle pulse formation via reflection of single-cycle pulse from linear medium \cite{Arkhipov-OL}. The idea is based on a coherent control of low-frequency oscillations \cite{Weiner} and free induction decay \cite{Allen} in a nonlinear medium excited by a femtosecond pulse train. The first pulse in the train switches the low-frequency oscillations on whereas the other stops it. Pulse duration and delay are smaller than medium polarization relaxation time $T_2$. Hence, pulses interact with medium coherently and medium polarization oscillations are created between pulses (free induction decay \cite{Allen}).  In high frequency range, similar idea was utilized in \cite{PakhomovPRA} for quasi-unipolar pulse generation. Our method possesses high efficiency with the pump pulses within the reach of current technologies. Importantly, we can control the shape of the emitted THz waveforms in a simple way, via duration and amplitude of the pump pulses as well as the thickness of the medium layer.

\begin{figure}[htpb]
\centering
\includegraphics[width=.9\linewidth]{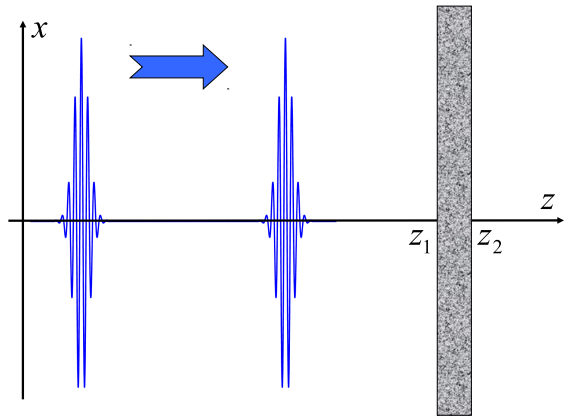}
\caption{(Color online) An optically thin layer of a nonlinear medium (shadowed area) is irradiated by a train of two few-cycle pulses (blue curves) having a plane wavefront and propagating along the $z$ axis, which is orthogonal to the layer. The time delay between pulses equals $T_0/2$. The pulses are linearly polarized along the $x$ axis.}
\label{fig1}
\end{figure}

Let us consider an optically thin plane layer of a nonlinear medium having resonances in THz range of thickness $h$, which is irradiated by a train of two few-cycle optical pump pulses as shown in Fig.~\ref{fig1}. The pump pulses are linearly polarized, have plane wavefronts and propagate along the $z$ axis with the velocity of light in vacuum $c$. Under these assumptions, we can describe our problem by the one-dimensional (1D) scalar wave equation for the electric field $E(z,t)$ coupled to the equation describing the response of the nonlinear medium  \cite{PakhomovPRA}:
\begin{eqnarray}
\frac{\partial^2 E(z,t)}{\partial z^2}-\frac{1}{c^2}\frac{\partial^2 E(z,t)}{\partial t^2} = \frac{4\pi}{c^2}  \frac{\partial^2 P(z,t)}{\partial t^2},
\label{eq_wave} \\
\ddot{P}+\nu\dot{P}+\omega_{0}^2P=g[E(t)] E(t),
\label{eq_medium} 
\end{eqnarray}
where  $t$ is time, $P(z,t)$ is the  polarization of the medium, $\omega_{0}$ and $\nu$ are its resonant frequency and decay rate of the medium. We assume $\omega_0$ in the THz frequency range. $g[E(t)]$ is the function  
describing nonlinear coupling of the low-frequency oscillators to the field; in our case we assume the lowest order approximation $g[E(t)] = g_0 E(t)$. This system of equations is written beyond the commonly used slowly-varying envelope (SVEA) and rotating wave approximations (RWA). Here we consider one-dimensional (1D) problem, which is  a good approximation within the propagation distances much smaller than the diffraction length $L_{d} = ka^2$, where $k=2\pi/\lambda$, $a$ is the beam transverse size, and $\lambda$ is the wavelength. Therefore one should use the beams so wide that for the most part (energetically) of our beam/pulse the requirement $L < L_{d}$ is satisfied.

Eq.~(\ref{eq_medium}) can be employed to describe the optical response of different media. In particular, low-frequency oscillators in a molecular Raman-active media (RAM)  \cite{Platonenko, Akhmanov} are governed by a similar equation due to their nonlinear bonding to high-frequency electron oscillations. Besides, Eq.~(\ref{eq_medium}) can be also expected to adequately describe the dynamics of nonlinearly coupled plasmon resonances in metallic nanoparticles, coupled semiconductor quantum dots or other hybrid optical materials \cite{Ginzburg}.

In a one-dimensional problem the field reflected from the layer is given as \cite{Arkhipov-OL}:
\begin{eqnarray}
E_{\mathrm{g}}(z,t)=-\frac{1}{2c\varepsilon_0} \int_{z_1}^{z_2} \frac{\partial P\Big( z',t-\frac{|z-z'|}{c}\Big )}{\partial t} dz'.
\label{Field_1D_thick} 
\end{eqnarray}
Eq.~(\ref{Field_1D_thick})  implies that the response of the medium  is proportional to the velocity of the charges instead of their acceleration (as it happens in a three-dimensional geometry). This fact is essential for the following consideration.

Let us suppose that the layer consists of the particles possessing at least one resonance in THz range with the frequency $\omega_0$, and the distance between the driver pulses in Fig.~\ref{fig1} is $T_0/2 = \pi/\omega_0$, that is, equal to the half of the period of the resonant oscillations in the medium. The electric field of the pair of such driving pulses is described by:
\begin{eqnarray}
\nonumber
E_{\mathrm{in}}(z,t)=E_0 e^{-(t-\frac{z}{c} )^2/\tau_{p}^2}\sin \Omega_p \Big(t-\frac{z}{c} \Big) + \\
E_0e^{-(t-\frac{z}{c}-T_0/2)^2/\tau_{p}^2}\sin \Omega_p \Big(t-\frac{z}{c}-T_0/2 \Big) ,
\label{pump} 
\end{eqnarray}
where $\Omega_p$ is the central frequency of the pump pulse (which we assume to be in the optical range) and $\tau_p$ is the duration of the pump pulses. 

The situation considered here is similar to well-known coherent control of molecular vibrations by femtosecond pulse train \cite{Weiner}. The pulse interacts with the film coherently, i.e. pulse duration and delay between pulses are assumed to be smaller than polarization relaxation relaxation time of the medium $T_2=1/\nu$. Hence, under these conditions the 1st pulse induces the  oscillations of polarization at the frequency $\omega_0$, which exist via time $T_2$  and the 2nd pulse stops the oscillations. Due to free induction decay  \cite{Allen} the oscillating dipoles radiate even without the external field, producing a THz waveform \cite{PakhomovSR}. 

As a first step, to make an estimation and get the feeling of the processes which take place during the system's evolution we simplify the model by suggesting that the polarization decay can be neglected. In this case, the general solution of Eq.~(\ref{eq_medium}) with the right-hand side Eq.~(\ref{pump}) is given by \cite{PakhomovPRA}: 
\begin{eqnarray}
\nonumber
\dot{P}(t) = \cos(\omega_0 t) \int_{-\infty}^t   g_0  E_{\mathrm{in}}(t')^2 \cos(\omega_0 t') dt'  +  \\
\sin(\omega_0 t) \int_{-\infty}^t  g_0 E_{\mathrm{in}}(t')^2 \sin(\omega_0 t') dt'.
\label{emission}
\end{eqnarray}
Suppose that the duration of the excitation pulses is much smaller than the period of the resonant optical transition in the medium: $\tau_p \ll T_0$, and the layer thickness is much smaller than the transition wavelength: $h \ll \lambda_0$. Given that we can divide the integration region in Eq.~(\ref{emission}) into three time intervals.
The first of them corresponds to the action of the first pump pulse. During this time the dipoles start to oscillate, emitting radiation. The emitted field, followed by the dipole dynamics, rapidly and monotonically changes from zero to the value given by:
\begin{equation}
\nonumber
-{E}_m = -\frac{1}{2c\varepsilon_0} \dot{P}_m h  \approx -\frac{1}{2c\varepsilon_0} \Pi h \cos \omega_0 \tau_p  \approx -\frac{1}{2c\varepsilon_0} \Pi h,
\label{Pi_max}
\end{equation}
with
\begin{equation}
\Pi = \int_{-\infty}^{+\infty} g_0 E_{\mathrm{in}}(t')^2 \cos(\omega_0 t') dt',
\label{eq:pval}
\end{equation}
where the integration is performed assuming only the first pump pulse is present.

At the second stage, in between the excitation pulses ($0 < t < T_0/2$) the  resonant oscillations lead to the emission due to free polarization decay:
\begin{eqnarray}
\nonumber
E(t) = -\frac{1}{2c\varepsilon_0} \dot{P}(t) h  \approx -\frac{1}{2c\varepsilon_0} \Pi h \cos \omega_0 t.
\label{P2}
\end{eqnarray}

Finally, the second pump pulse is chosen in such a way that the oscillators are stopped and thus, during the action of the second pulse, the emitted field is monotonically reduced from 
\begin{eqnarray}
\nonumber
E_m \approx \frac{1}{2c\varepsilon_0} \Pi h
\label{P3}
\end{eqnarray}
to zero. As a result,  a single-cycle terahertz waveform, or, to be more specific, a single sign-changing oscillation of the electric field is produced (see Visualization 1).

The simple theory presented above is to give only a rough estimation of the generation dynamics. 
More generally, the exact waveshape obtained at reflection depends on the ratio of the pump pulse durations $\tau_p$ and the 
time delay between the pump pulses. 
If $\tau_p \ll T_0$, the reflected pulse will possess abrupt jumps of the electric field amplitude at the 
edges of the pulse. These sharp edges can be smoothed by the increase  $\tau_p$.

The aforementioned influence of the parameters of the pump pulses on the generated waveform was calculated and analyzed numerically by solving the system of Eqs.~(\ref{eq_wave})-(\ref{eq_medium}). We have applied time-domain finite-difference method (FDTD) to solve the wave equation Eq.~(\ref{eq_wave}) and 4-th order Runge-Kutta method for the equation for the medium polarization Eq.~(\ref{eq_medium}).

\begin{figure}[htpb]
\centering
\includegraphics[width=1\linewidth]{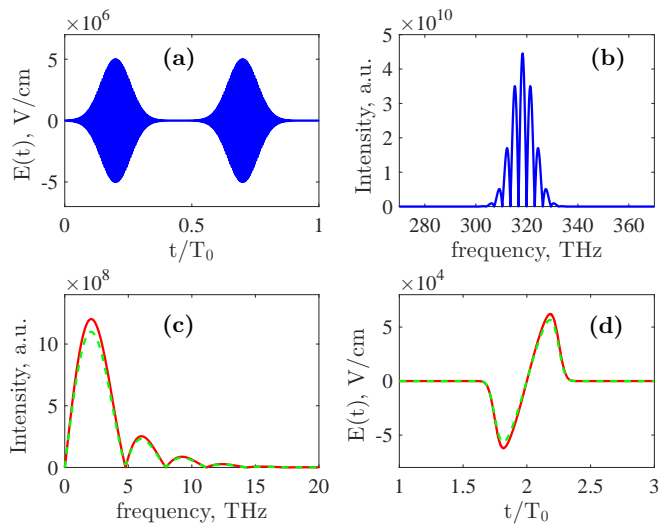}
\caption{(Color online) The driving field (a), its spectrum (b); the spectra (c) and the temporal profiles (d) of the single-cycle THz pulses generated from a layer governed by Eq.~(\ref{eq_medium}) (red curves) and from 
a molecular RAM layer (green curves) described by  Eqs.~(\ref{eqx}),(\ref{eqy}) after filtering out the high-frequency components. 
The parameters are listed in the text. }
\label{fig23}
\end{figure}

Fig.~\ref{fig23} shows the pump pulses (a), their spectra (b) as well as an example of the  single-cycle THz pulse (d,c) (red curves), obtained as a result of reflection from the layer. In this example, the central frequency of the pump pulse is $\Omega_p=2\cdot 10^{15}$ rad/s and the medium  resonant frequency is $\omega_0=10^{13}$ rad/s ($T_0=6.28\cdot10^{-13}$ s). We take in our simulations $g_0 = 5\cdot10^{5}$ m$^2$/(C $\cdot$ s$^2$) in order to meet the typical values of the parameters of RAM. 
Pump field amplitude amounts $E_0=5\cdot10^6$ V/cm (intensity $I_0=3.3\cdot10^{10}$ W/cm$^2$), pulse duration is $\tau_p = 50$ fs and the dephasing time $1/\nu$ is supposed to be large enough to be neglected. The particle density in the layer was taken $N_0=5\cdot10^{21}$ cm$^{-3}$ and medium length was chosen to be $h=10$ $\mu$m.

Fig.~\ref{fig23}c,d clearly show that the generated field is indeed a single-cycle THz waveform. Its amplitude is $E_{g0} \approx 6 \cdot10^{4}$ V/cm, what gives the conversion efficiency:
\begin{eqnarray}
\eta = \frac{ \int_{-\infty}^{+\infty} E_{g}^2(t) dt } {\int_{-\infty}^{+\infty} E_{in}^2(t) dt} \approx 3\cdot10^{-4}.
\label{eta}
\end{eqnarray}

We now aim to illustrate the findings above, obtained for the general model Eq.~(\ref{eq_medium}), using more specific model of a molecular RAM consisting of two nonlinearly-coupled harmonic oscillators - high-frequency (electrons) oscillator (HFO) and low-frequency (nucleus) one (LFO)  \cite{Platonenko,Akhmanov}. As an example of RAM one can consider thin films of solid state materials having isolated vibrational resonances in THz range. Those can be for instance molecular crystals \cite{Weiner}, which are actively used in the experiments on coherent control.  If we denote the normal coordinate of LFO oscillations $y$ and the normal coordinate of HFO oscillations $x$, then the dynamics of RAM under pulsed excitation is governed by the following equations \cite{Platonenko,Akhmanov}
\begin{eqnarray}
\ddot{x}+\Gamma_e \dot{x} + \Omega_{0}^2 x =& \frac{q}{m}E_{in}(t) - \frac{\gamma}{m}xy,
\label{eqx} \\
\ddot{y}+\Gamma_n \dot{y} + \omega_{0}^2 y =& -\frac{\gamma}{2M}x^2.
\label{eqy}
\end{eqnarray}
Here $M$ and $m$ are the effective masses of the LFO and HFO, factor $\gamma$ stands for the strength of the nonlinear bonding between HFO and LFO, $\Gamma_e$ and $\Gamma_n$ are the damping rates of HFO and LFO respectively, $\omega_0=2\pi/T_0$ is the resonant frequency of LFO, $\Omega_0=2\pi/T_{\Omega}$ is the resonant frequency of the HFO. We assume here, that the pumping frequency is far from the vibrational resonances of RAM (LFO resonances) and thus the population dynamics can be reliably neglected.

Field emitted by the RAM layer is calculated using Eq.~(\ref{Field_1D_thick}), where we take $P = qN_0(x+y)$, with the density of RAM oscillators $N_0$ and electric charge $q$ (Fig.~\ref{fig23}a). 
The parameters of the medium are: $\omega_0 = 10^{13}$ rad/s, $\Omega_0 = 10^{15}$ rad/s, $m = 9.1\cdot10^{-31}$ kg (electron mass), $M = 3\cdot10^{-26}$ kg (nucleus mass), electric charge $q = 1.6\cdot10^{-19}$ C, $N_0=5\cdot10^{21}$ cm$^{-3}$, $\gamma = 10^{10}$ J/m$^3$, $\Gamma_e = 10^{14}$ s$^{-1}$, $\Gamma_n = 0$ s$^{-1}$, $E_{0} = 5\cdot10^{6}$ V/cm, $\tau_p = 50$ fs, $\Omega_p = 2\cdot10^{15}$ rad/s.

The spectrum of the produced pulse consists of two well-separated parts, arising from the oscillations of HFO and  LFO. To obtain low-frequency THz response we remove the high frequency components above $\omega_C = 5\cdot10^{14}$ rad/s by a low-pass filter in the form of the Heaviside step-function $H(\omega) = \Theta(\omega_C - \omega)$. The resulting spectrum and waveshape  are shown in Fig.~\ref{fig23}c,d  (green curves). 
One can see, that the pulse obtained with the LFO-HFO model coincides with the one obtained using the model Eq.~(\ref{eq_medium}) (Fig.~\ref{fig23}d). 

\begin{figure}[htpb]
\centering
\includegraphics[width=.9\linewidth]{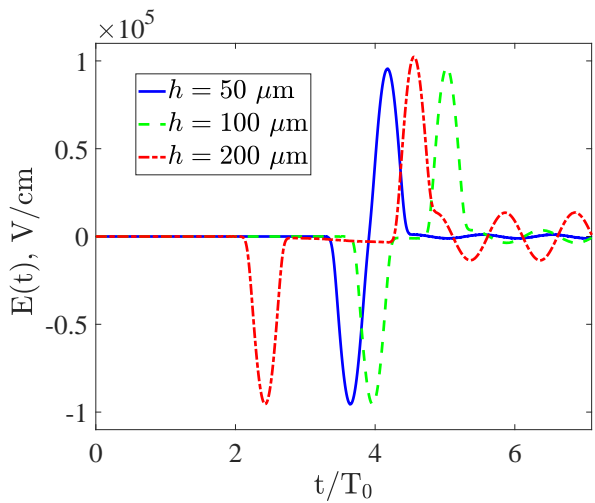}
\caption{(Color online) THz pulses generated for different values of the thickness $h$ of the layer. Other parameters are the same as in Fig.~\ref{fig23}.}
\label{fig4}
\end{figure}

As a next step, we consider how the THz waveshape  changes with the  parameters of the driving pulses. In order to do that we performed a series of numerical simulations of Eqs.~(\ref{eq_wave})-(\ref{eq_medium}) for different layer thicknesses $h$ and pump field durations $\tau_p$. The generated pulse shapes for different $h$ are plotted in Fig.~\ref{fig4}. It is seen, that as the thickness of the layer increases, the pulse shape starts to deviate from a single THz cycle. 
It stems from the fact that during the propagation the temporal shape of the excitation femtosecond pulses changes noticeably.
As a result, the second pump pulse does not  stop anymore the low-frequency oscillators as it happens in a thin layer. Hence, a long oscillating tail arises, see Fig.~\ref{fig4}. Nevertheless, the conversion efficiency Eq.~(\ref{eta}) can be somewhat increased with the layer thickness $h$. For instance, for $h=50$ $\mu$m,  $\eta \approx 10^{-3}$ is achieved.

\begin{figure}[htpb]
\centering
\includegraphics[width=.9\linewidth]{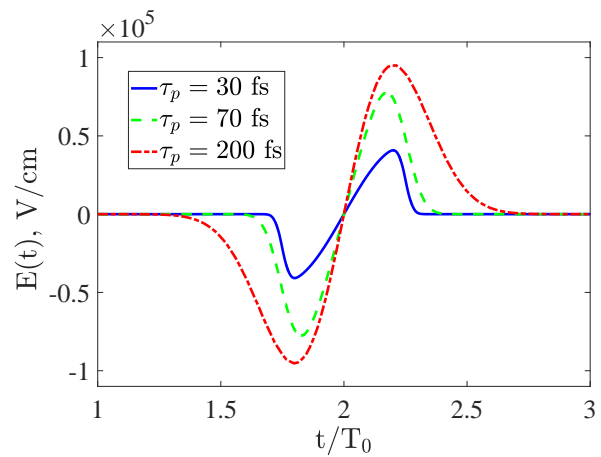}
\caption{(Color online) Generated THz pulses for different values of the duration of the pump pulses $\tau_p$. Other parameters are the same as in Fig.~\ref{fig23}.}
\label{fig5}
\end{figure}

Furthermore, the dependence on the input pulse duration is also important. Since the polarization induced in the layer is proportional to the pulse energy as it is seen in Eq.~(\ref{eq:pval}), 
larger durations of pump pulses lead also to higher energies of the response of the layer \cite{PakhomovPRA}. 
Therefore, it is expected that the 
THz pulse energy can be boosted with the increase of the pump pulse duration $\tau_p$. 
This fact is illustrated by Fig.~\ref{fig5}, where the temporal profile of the emitted field for different values of $\tau_p$ is plotted. 
We remark that in addition to the dependencies we considered up to now the amplitude of the THz waves depends on the the amplitude of the pump pulses, their central frequency as well as the particle density $N_0$.

In conclusion, we have proposed and studied theoretically a new method for generation of tunable  single-cycle THz waves via reflection of a train of few-cycle femtosecond pulses from a thin nonlinear layer of a medium which possesses resonances in THz range. The one-dimensional propagation problem was considered, which is valid in a single-mode fiber or at distances much smaller than the diffraction length. The method is based on the free induction decay and coherent control of the low-frequency oscillations arising in the medium using femtosecond pulse trains and also utilized peculiarities of nonlinear reflection in 1D case.

We have demonstrated that our method can exhibit reasonable conversion efficiency up to $\sim 10^{-3}$.
The power and waveshape of the generated THz pulse can be controlled by the excitation pulse energy and duration, as well as by the thickness of the layer. Other advantages of the proposed method are its simplicity and compactness. It can be realized in a broad class of nonlinear RAM layers with commercially available femtosecond pulse sources.  Thus, we believe that the proposed method can serve as a novel high-efficient approach to generate THz waveforms with a tunable waveshape.

\section*{Funding Information}
R.A. and A.P. thank Russian Science Foundation (project 19-72-00012). I.B. thanks Deutsche Forschungsgemeinschaft (DFG, German Research Foundation), projects BA 4156/4-2 and MO 850-19/2,
as well as Germany’s Excellence Strategy within the Cluster of Excellence PhoenixD (EXC 2122, Project ID 390833453).

\bibliography{optics}

\begin{thebibliography}{10}
\newcommand{\enquote}[1]{#1}
\bibitem{Roskos}
H. G. Roskos, M. D. Thomson, M. Kress, T. L{\"o}ffler, "Broadband THz emission from gas plasmas induced by femtosecond optical pulses: From fundamentals to applications," Laser  Photon. Rev. {\bf 1}, 349--368 (2007).

\bibitem{Reiman} 
K. Reiman, "Table-top sources of ultrashort THz pulses," Rep. Progr. Phys. \textbf{70}, 1597--1632 (2007).
 
 \bibitem{Lepeshov}
S. Lepeshov, A. Gorodetsky, A. Krasnok, E. Rafailov, P. Belov, "Enhancement of terahertz photoconductive antenna operation by optical nanoantennas," Laser Photon. Rev. \textbf{11}, 1770001 (2016).

\bibitem{Kim1}
K. Kim, J. Glownia, A. Taylor, and G. Rodriguez, "Terahertz emission from ultrafast ionizing air in symmetry-broken laser fields," Opt. Expr. \textbf{15}, 4577--4584 (2007).


\bibitem{Kress1}
M. Kress, T. L{\"o}ffler, M. D. Thomson, R. D{\"o}rner, H. Gimpel, K. Zrost, T. Ergler, R. Moshammer, U. Morgner, J. Ullrich, H. G. Roskos, "Determination of the carrier-envelope phase of few-cycle laser pulses with terahertz-emission spectroscopy," Nature Phys. \textbf{2}, 327--331 (2006).

\bibitem{Babushkin1} 
I. Babushkin, S. Skupin, A. Husakou, C. K{\"o}hler, E. Cabrera-Granado, L. Bergé, and J .Herrmann, "Tailoring terahertz radiation by controlling tunnel photoionization events in gases," New J. Phys. \textbf{13}, 123029 (2011).

\bibitem{Tcypkin}
 A. N. Tcypkin, E. A. Ponomareva, S. E. Putilin, S. V.  Smirnov, S. A. Shtumpf, M. V.  Melnik, E. Yiwen, S.A. Kozlov, X. C. Zhang, "Flat liquid jet as a highly efficient source of terahertz radiation",  Opt. Expr. \textbf{27(11)}, 15485--15494 (2019). 

\bibitem{Berry2}
C. Berry, N. Wang, M. Hashemi, M. Unlu, and M. Jarrahi, "Significant performance enhancement in photoconductive terahertz optoelectronics by incorporating plasmonic contact electrodes," Nature Commun. \textbf{4}, 1622 (2013).

\bibitem{Gorodetsky}
S. Lepeshov, A. Gorodetsky, A. Krasnok, N. Toropov, T. A. Vartanyan, P. Belov, A. Alú, and E. U. Rafailov, "Boosting Terahertz Photoconductive Antenna Performance with Optimised Plasmonic Nanostructures,"  Sci. Rep. \textbf{8}, 6624 (2018).

\bibitem{Palfalvi}
L. Pálfalvi, G. Tóth,  L. Tokodi,  Z. Márton, J. A. Fülöp, G. Almási, J. Hebling, "Numerical investigation of a scalable setup for efficient terahertz generation using a segmented tilted-pulse-front excitation", Opt. Expr. \textbf{25(24)}, 29560--29573 (2017).

\bibitem{Nugraha}
P. S. Nugraha, G. Krizsán, C. Lombosi, L. Pálfalvi, G. Tóth, G. Almási, J. A. F{\"u}l{\"o}p, and J. Hebling, "Demonstration of a tilted-pulse-front pumped plane-parallel slab terahertz source", Opt. Lett. \textbf{44(4)}, 1023--1026 (2019).

\bibitem{Toth}
G. Tóth, L. Pálfalvi, J. A. Fülöp, G. Krizsán, N. H. Matlis, G. Almási, and J. Hebling, “Numerical investigation of imaging-free terahertz generation setup using segmented tilted-pulse-front excitation,” Opt. Expr. \textbf{27}, 7762-7776 (2019).

\bibitem{Arkhipov-OL}
M. V. Arkhipov, R. M. Arkhipov,  A. V. Pakhomov, I. V. Babushkin, A. Demircan, U. Morgner, and  N. N. Rosanov, "Generation of unipolar half-cycle pulses via unusual reflection of a single-cycle pulse from an optically thin metallic or dielectric layer," Opt. Lett. \textbf{42}, 2189--2192 (2017).

\bibitem{Weiner}
A. M. Weiner, D. E. Leaird,  G. P. Wiederrecht, K. A. Nelson,  "Femtosecond multiple-pulse impulsive stimulated Raman scattering spectroscopy," J. Opt. Soc. Am. B \textbf{8(6)}, 1264--1275 (1991).
 
\bibitem{Allen}
L. Allen, J. H. Eberly, \textit{Optical Resonance and Two-Level Atoms} (Dover Publications, New York, 1987).  

\bibitem{PakhomovPRA}
A. V. Pakhomov,  R. M. Arkhipov,  I. V. Babushkin,  M. V. Arkhipov, Yu. A. Tolmachev, N. N. Rosanov, "All-optical control of unipolar pulse generation in a resonant medium with nonlinear field coupling," Phys. Rev. A \textbf{95}, 013804 (2017).

\bibitem{Platonenko}
V. T. Platonenko, R. V. Khokhlov, "On the Mechanism of Operation of a Raman Laser," Sov. Phys. JETP \textbf{19}, 378--381 (1964).

\bibitem{Akhmanov}
S. A. Akhmanov, S. Yu. Nikitin, \textit{Physical Optics} (Nauka, Moscow, 2004; Clarendon, Oxford, 1997).

\bibitem{Ginzburg}
P. Ginzburg, A. Krasavin, Y. Sonnefraud, A. Murphy, R. J. Pollard, S. A. Maier, and A. V. Zayats, "Nonlinearly coupled localized plasmon resonances: Resonant second-harmonic generation," Phys. Rev. B \textbf{86}, 085422 (2012).

\bibitem{PakhomovSR}
A. V. Pakhomov, R. M. Arkhipov, M. V. Arkhipov, A. Demircan, U. Morgner, N. N. Rosanov, I. V. Babushkin, "Unusual terahertz waveforms from a resonant medium controlled by diffractive optical elements," Sci. Rep. \textbf{9}, 7444 (2019). 




\end{thebibliography}

\end{document}